\documentclass[aps,twocolumn,showpacs,amsmath,amssymb,floatfix,eqsecnum,superscriptaddress]{revtex4}
\usepackage{graphicx}
\usepackage{dcolumn}
\usepackage{bm}
\usepackage{epsfig,epsf}
\usepackage{epstopdf}
\pdfoutput=1

\newcommand{\beq}{\begin{eqnarray}}
\newcommand{\eeq}{\end{eqnarray}}
\def\be{\begin{equation}}
\def\ee{\end{equation}}
\newcommand\eq[1]{Eq.~(\ref{#1})}

\usepackage{graphicx}
\usepackage{amssymb}
\usepackage{epstopdf}
\usepackage{color}
\usepackage{hyperref}
\usepackage{amsmath}
\usepackage{bbold}

\begin{document}

\title{Exact expressions for the partition function of the one-dimensional Ising model in the fixed-$M$ ensemble.}

\author{Daniel Dantchev}
\affiliation{Institute of Mechanics--Bulgarian Academy of Sciences, Academic Georgy Bonchev St. Building 4, 1113, Sofia, Bulgaria}
\affiliation{Department of Physics and Astronomy, University of California, Los Angeles, CA 90095}
\affiliation{Max-Planck-Institut f\"{u}r Intelligente Systeme, Heisenbergstrasse 3, D-70569 Stuttgart, Germany} 
\affiliation{
	IV. Institut f\"{u}r Theoretische Physik, Universit\"{a}t Stuttgart, Pfaffenwaldring 57, D-70569 Stuttgart, Germany}
\author{Joseph Rudnick}
\affiliation{Department of Physics and Astronomy, University of California, Los Angeles, CA 90095}

\date{\today}

\begin{abstract}
We obtain exact closed-form expressions for the partition function of the one-dimensional Ising model in the fixed-$M$ ensemble, for three commonly-used boundary conditions: periodic, antiperiodic and Dirichlet. These expressions allow for the determination of fluctuation-induced forces in the canonical ensemble, which we term Helmholtz forces. The thermodynamic expressions and the calculations flowing from them should provide insights into the nature and behavior of fluctuation induced forces in interesting and as-yet unexplored regimes. 
\end{abstract}

\maketitle

\section{Introduction} \label{sec:intro}

Except for the ideal gas, the Ising model is probably the most well-known model in statistical physics. Its solution for the one dimensional case with a temperature $T$ and external field $h$---and for the two-dimensional case in zero external field---can be found in most textbooks of statistical mechanics---see, e.g., \cite{Baxter1982,Huang1987,Pathria2011,Berlinsky2019}.  Inspection of the literature shows, perhaps surprisingly, that the solution of the Ising model in one or any higher dimension is not available when the magnetization $M$ is fixed. In this article we fill that gap and provide closed-form expressions for the cases  of the finite Ising chain with fixed total magnetization $M$ under periodic, antiperiodic and free boundary conditions. The resulting expressions, which involve hypergeometric functions, differ non-trivially from corresponding expressions in the fixed-$h$ ensemble. 

The lack of expressions for the one-dimensional Ising model with fixed magnetization $M$ is not due to a lack of interest into the problem. We stress, as noted in \cite{Henkel}, that in  customarily  considered
applications of the equilibrium Ising model to binary alloys or binary liquids, if one insist on full rigor,  the case with $M$ fixed must be addressed. Thus, there have been attempts to solve the problem of Ising chain with fixed $M$ \cite{Henkel,Wang2017,Racz}.  In Refs. \cite{Henkel,Wang2017} it is attacked via the transfer matrix method.  Ref. \cite{Henkel} reports success in deriving a closed form expression for the partition function for the case $M=0$ under periodic boundary conditions, when the chain contains an even number of spins.   In Ref. \cite{Wang2017} the focus is on the asymptotic behavior of the free energy of the Ising chain with $M=0$ and under periodic boundary conditions and chain length $N\gg 1$ in the temperature regime when the correlation length of the chain $\xi$ is kept finite, i.e., excluding the regime $T\to0$. In Ref. \cite{Racz} the finite-size scaling functions for the probability distribution of the
magnetization in the one-dimensional Ising model has been investigated. The  functions are evaluated in the limit $T\to 0$ and $N\to\infty$ with $N/\xi$ kept
finite. Exact results for periodic, antiperiodic, free, and block boundary conditions have been obtained. The approach used there is based on a combinatorial approach to counting the domains of up and down spins. This is similar to the approach we use for our study of a fully finite Ising chain with periodic, antiperiodic and free boundary conditions. 

Knowledge of the partition function leads to the calculation of the Helmholtz free energy, which allows for the determination of a fluctuation-induced force in the fixed $M$-ensemble.  This can be achieved in a manner similar to the derivation of the Casimir force for critical systems in the grand-canonical $T$-$h$ ensemble: 
\begin{equation}
	\label{CasDef}
	\beta F_{\rm Cas}^{(\zeta)}(T,h,L)\equiv- \frac{\partial}{\partial L}f_{\rm ex}^{(\zeta)}(T,h,L)
\end{equation}
where
\begin{equation}
	\label{excess_free_energy_definition}
	f_{\rm ex}^{(\zeta)}(T,h,L) \equiv f^{(\zeta)}(T,h,L)-L f_b(T,h)
\end{equation}
is the so-called excess (over the bulk) free energy per area and per $k_B T$.
Here one envisages a system in film geometry $\infty^{d-1}\times L$, $L\equiv L_\perp$, with boundary conditions $\zeta$ imposed along the spatial direction of finite extent $L$, and with total free energy ${\cal F}_{ {\rm tot}}^{(\zeta)}$.
Here   $f^{(\zeta)}(T,h,L)\equiv \lim_{A\to\infty}{\cal F}_{ {\rm tot}}^{(\zeta)}/A$  is the free energy per area $A$ of the system. 
Along these lines we define 
\begin{equation}
	\label{HelmDef}
	\beta F_{\rm H}^{(\zeta)}(T,M,L)\equiv- \frac{\partial}{\partial L}f_{\rm ex}^{(\zeta)}(T,M,L)
\end{equation}
where
\begin{equation}
	\label{excess_free_energy_definition_M}
	f_{\rm ex}^{(\zeta)}(T,M,L) \equiv f^{(\zeta)}(T,M,L)-L f_H(T,m),
\end{equation}
with $m=\lim_{L, A\to \infty}M/(LA)$, and $f_H$ is the Helmholtz free energy density of the ``bulk'' system. 
We will show that the so-defined \textit{Helmholtz fluctuation induced force } has a behavior very different from that one of the Casimir force. Explicitly, we will demonstrate that for the Ising chain with fixed $M$ under periodic boundary conditions $F_{\rm H}^{\rm (per)}(T,M,L)$ can, depending on the temperature $T$, be attractive or repulsive, while $ F_{\rm Cas}^{\rm (per)}(T,h,L)$ is only attractive. We note that the issue of the ensemble dependence of fluctuation induced forces pertinent to the ensemble has not, to our knowledge, been studied up to now. This issue is by no means limited to the Ising chain and can be addressed, in principle, in any model of interest. The analysis reported here can also be viewed as a useful addition to approaches to fluctuation-induced forces in the fixed-$M$ ensemble based on Ginzburg-Landau-Wilson Hamiltonians \cite{Dietrich1,Dietrich2,Dietrich3} in which one studied the usual Casimir force.

Before turning to the specific calculations related to the Ising chain with fixed magnetization we note that recently one dimensional and quasi one-dimensional systems have been the objects of intensified experimental interest---see, e.g., \cite{vdW2022} and references cited therein. Some of these, like ${\rm TaSe_3}$, are
quasi-one dimensional in the sense that they have strong covalent bonds in one
direction along the atomic chains and weaker bonds in the perpendicular
plane \cite{Stolyarov2016}. Others are
more properly considered true one dimensional materials, in that they have covalent bonds only
along the atomic chains and much weaker van der Waals interactions
in perpendicular directions \cite{BALANDIN2022}. One-dimensional van der Waals materials have emerged as an entirely new field, which encompasses interdisciplinary work by physicists, chemists, materials
scientists, and engineers \cite{vdW2022}. The Ising chain considered here can be seen as the simplest possible example of such a one-dimensional material. Earlier experimental realization of one-dimensional Ising model have been considered in Refs. \cite{Armitage1,Armitage2,Armitage3,Armitage4,vdW2022}. The one-dimensional Ising model in a transverse field has proven an important  experimental realization of a system with a quantum phase transition \cite{Armitage4}. 

The Ising chain in $T$-$h$  ensemble manifests scaling behavior in the vicinity of its zero-temperature ordered state, and it is a test-bed for exploring the influence of finite-size scaling on critical properties, including the connection between the behavior of fluctuation-induced forces in the critical regime and scaling predictions \cite{rudzanabshak2010, Dantchev2022}.

Below, we describe the derivation of the exact results for the partition function of an Ising chain with fixed magnetization $M$ under different boundary conditions. In what follows we will assume a lattice constant $a=1$, so that instead of $L$ we will use $N$ as a measure of the length of the chain.

\section{Ising chain with fixed $M$: the combinatorics of domains} \label{sec:periodicdomains}

As in the case of all boundary conditions considered here, the partition function to be evaluated is the sum over spin states of the Boltzmann factor
\begin{equation}
\exp\left[K \sum_{i=1}^{N-1}s_i s_{i+1}\right]  \label{eq:bf}
\end{equation}
where each spin variable takes on the values $\pm 1$. Fixing the total magnetism amounts to the constraint that the difference between the number of up spins, $N_+$ and the number of down spins, $N_-$ is equal to $M$.

The key step in the calculation of the partition function is the determination of the number of ways in which the spins can arrange themselves into alternating spin-up and spin-down domains, subject to the requirement of a fixed value of the total magnetization, $M$.  We start with equations that express the relationships between $M$, the number of up spins, $N_+$ and the number of down spins, $N_-$, along with the total number of spins, $N$:
 \begin{equation}
 	N = N_+ + N_-, \, \qquad \mbox{and}\qquad 
 	M =  N_+-N_- \label{eq:cf1}.
 \end{equation}
 Inverting these equations we find
 \begin{equation}
 	N_+  =  \frac{N+M}{2} \, \qquad \mbox{and}\qquad 
 	N_-  =  \frac{N-M}{2}. \label{eq:cf2}
 \end{equation}
 
\begin{figure}[htbp]
 		\includegraphics[width=3in]{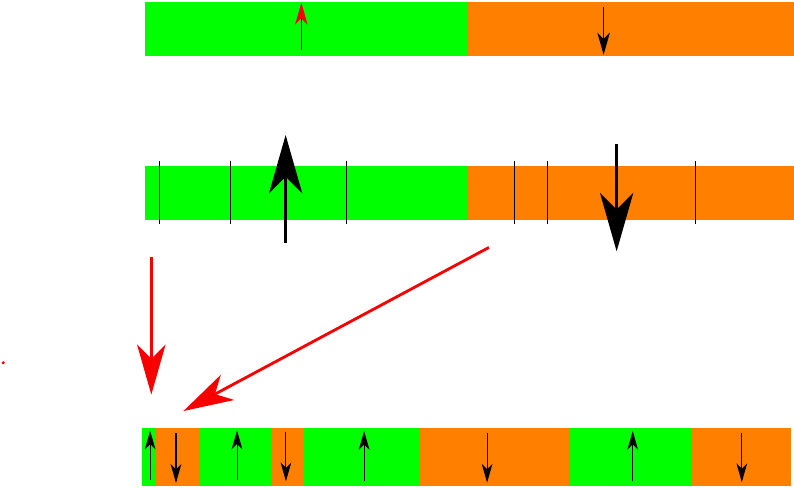}
 		\caption{Top portion: the domain of up spins (in green) and of down spins (in orange); middle portion: the domains are each divided into four smaller domains; bottom portion: the smaller domains are now interspersed; see the red arrows.}
 		\label{fig:steps}
 \end{figure}

For insight into the determination of domain statistics, we look at, say, the fourth leading contribution in an expansion of the partition function in powers of $\exp[-K]$. We start with a domain of $N_+$ up spins. We then partition that domain into four smaller domains. We do this by inserting three ``slices'' into the domain, effectively three walls between adjacent spins. 
 We then partition a domain of $N_-$ down spins into four smaller domains, which we insert between the domains of up spins. The process is illustrated in Fig. \ref{fig:steps}.

 We now calculate how many ways there are of subdividing each domain into four subdomains. In the case of the spin up domain, that quantity is 
 \begin{equation}
 	(N_+-1)(N_+-2)(N_+-3)/3! \label{eq:cf5}
 \end{equation}
 which  is the  number of ways of inserting three partitions between adjacent spins in a linear array of $N_+$ up spins. A similar expression holds for the number of ways of subdividing the domain of down spins. Making use of the relations  (\ref{eq:cf2}) and multiplying resulting expressions to obtain the number of ways of subdividing both domains we end up with the factor 
  \begin{equation}
 	((N-2)^2-M^2)((N-4)^2-M^2)((N-6)^2-M^2)/(4^3(3!)^2). \label{eq:cf9}
 \end{equation}
 
 We now join the ends of the set of domains up so that they form a ring, consistent with periodic boundary conditions, and we rotate the ring around to find out how many ways we can arrange the subdomains. This yields a factor of $N$. However, because we take all possible lengths for the set of subdomains we are overcounting by a factor of four, the number of pairs of domains. 
 The overall factor is thus
 \begin{equation}
 	\frac{N}{4}\frac{((N-2)^2-M^2)((N-4)^2-M^2)((N-6)^2-M^2)}{  4^3 (3!)^2}. \label{eq:cf10}
 \end{equation}
 To obtain the complete expression, we multiply the above by $\exp(-16K)$, corresponding to the energy cost of the eight  walls between the eight domains of the periodically continued array in Fig. \ref{fig:steps}. 
 
 In the general case of $2k$ alternating domains, the first factor of 4 in the denominator of (\ref{eq:cf10}) is replaced by $k$, while the two other factors become $4^{k-1}$ and $(k-1)!^2$. Thus, the general form of the denominator is
 \begin{equation}
 	\label{eq:gen-form-denomnator}
 4^{k-1} k ((k-1)!)^2.
 \end{equation}
Then, for the numerator one has
\begin{equation}
	\label{eq:gen-form-nominator}
	N\prod _{p=1}^{k-1} \left((N-2 p)^2-M^2\right).
\end{equation}
Taking into account that the energy of a configuration with $2k$ domains is $\exp [K (N-4 k)]$, for the contribution of this configuration in the statistical sum one obtains 
\begin{eqnarray}
	\label{eq:stat-sum-expansion}
	\lefteqn{\mathrm{Zterm}(N,M,K,k)} \\&=& \frac{N \exp (K (N-4 k)) \prod _{p=1}^{k-1} \left((N-2 p)^2-M^2\right)}{4^{k-1} k ((k-1)!)^2}.\nonumber 
\end{eqnarray}
The form of the right hand side of (\ref{eq:stat-sum-expansion}) allows us to sum over $k$ from 0 to $\infty$ to obtain the partition function $Z^{({\rm per})}(N,M,K)$. The result is a closed-form expression that is exact when $N$ and $M$ are both even or odd integers with $|M|<N$,  and that smoothly interpolates between the exact expression for all other values of $N$ and $M$ with $|M|<N$. The case $|M|=N$ is exceptional, but trivial to determine. The result is 
\begin{eqnarray}
	\label{eq:statistical-sum}
&&	Z^{(\rm per)}(N,K,M) = N e^{K (N-4)}\\&& \times \,_2F_1\left(\frac{1}{2} (-M-N+2),\frac{1}{2} (M-N+2);2;e^{-4 K}\right), \nonumber
\end{eqnarray}
where $_2F_1(\alpha,\beta;\gamma;z)$ is the generalized hypergeometric function \cite{A&S1970}.

Similar calculations \cite{supmat} lead to expressions for the partition functions in the case of antiperiodic and Dirichlet boundary conditions. 
When the boundary conditions are Dirichlet, we have
\begin{widetext}
\begin{eqnarray}
	Z^{(D)}(N,K,M) 
	& = & e^{K (N-1)} \Bigg[2 e^{-2 K} \, _2F_1\left(\frac{1}{2} (-M-N+2),\frac{1}{2}
	(M-N+2);1;e^{-4 K}\right) \nonumber \\ &&-\frac{1}{2} e^{-4 K} (M-N+2) \, _2F_1\left(\frac{1}{2}
	(-M-N+2),\frac{1}{2} (M-N+4);2;e^{-4 K}\right) \nonumber \\ && + \frac{1}{2} e^{-4 K} (M+N-2) \,
	_2F_1\left(\frac{1}{2} (-M-N+4),\frac{1}{2} (M-N+2);2;e^{-4 K}\right)\Bigg] \nonumber \\  \label{eq:DDbc-main-text-appendix}
\end{eqnarray}
\end{widetext}
and when the boundary conditions are antiperiodic, the partition function is given by 
\begin{widetext}
\begin{eqnarray}
Z^{(\rm anti)}(N,K,M)
& = &e^{K (N-6)} \left[2 \left(e^{4 K}-1\right) \, _2F_1\left(\frac{1}{2} (-M-N+2),\frac{1}{2}
   (M-N+2);1;e^{-4 K}\right) \right. \nonumber \\ & & \left. +N \, _2F_1\left(\frac{1}{2} (-M-N+2),\frac{1}{2}
   (M-N+2);2;e^{-4 K}\right)\right]  \label{eq:cf33}
\end{eqnarray}
\end{widetext}
As in the case of periodic boundary conditions, the expressions above for the partition function when boundary conditions are Dirichlet and antiperiodic are exact except in the case of perfect alignment of the spins, when $M= \pm N$. 

If we write $M=mN$ and focus on the case $N \gg 1$, then the exact expressions above approach different forms. In the case of periodic boundary conditions, the partition function becomes
\begin{equation}
Z^{(\rm per)}_{\rm \lim}(N,K,m) = \frac{2}{N}\frac{e^{NK}x_t}{\sqrt{1-m^2}}I_1(x_t \sqrt{1-m^2}) \label{eq:scalingform}
\end{equation}
where $I_1$ is the modified Bessel function of order 1,  $x_t=Ne^{-2K}$ is the scaling combination $N/\xi_t$, $\xi_t $ being the correlation length \cite{Baxter1982} in the vicinity of the zero temperature critical point. This allows us to explore the scaling behavior of thermodynamic quantities close to $T=0$. Limiting forms for the antiperiodic and Dirichlet partition functions can also be obtained. 

The explicit formulas (\ref{eq:statistical-sum})--(\ref{eq:cf33}) allow one to obtain expressions involving derivatives with respect to the size, $N$, and the total magnetization, $M$, of the Ising chain. This is useful in the calculation of fluctuation-induced forces in the one-dimensional Ising system. Because of the nature of the ensemble in which the forces are generated, we refer to them as \textit{Helmholtz} forces. The determination of this kind of force requires that we specify precisely what is held constant in the finite Ising strip (the ``film'') and the infinite Ising system that borders it (the ``bulk''). Three of the possibilities are 1) constant total magnetization, $M$, 2) constant magnetization per site, $m=M/N$, and 3) constant number of up-spins, $N_+$. The last is relevant to lattice gas models. Consideration of the model  lead to the following observations.

\begin{enumerate}
\item  In the case of periodic and antiperiodic boundary conditions the degeneracy in the position of the borders between the domains with respect to translation results in a contribution to the Helmholtz free energy that is  logarithmic in $N$.  The implies lack of a perfect scaling. 
\item When $m$ in the fixed $m$ ensemble is not equal to $\pm1$,  the interfacial energy between the domains with coexisting phases plays a key role in the statistical mechanics of the system.  
\end{enumerate}

It is well known that in the grand canonical ensemble, i.e., fixed $h$, the Gibbs free energy of the finite system approaches the bulk limit \textit{exponentially} in $N$ (i.e. as $e^{- \alpha N}$ with $\alpha>0$)  as $N\to \infty$ for periodic boundary conditions. The properties  listed above imply that in systems with fixed $m$ the Helmholtz free energy possesses non-scaling contributions that vanish significantly more slowly than this exponential approach to the bulk behavior. 

\begin{figure}[htbp]
	\includegraphics[width=\columnwidth]{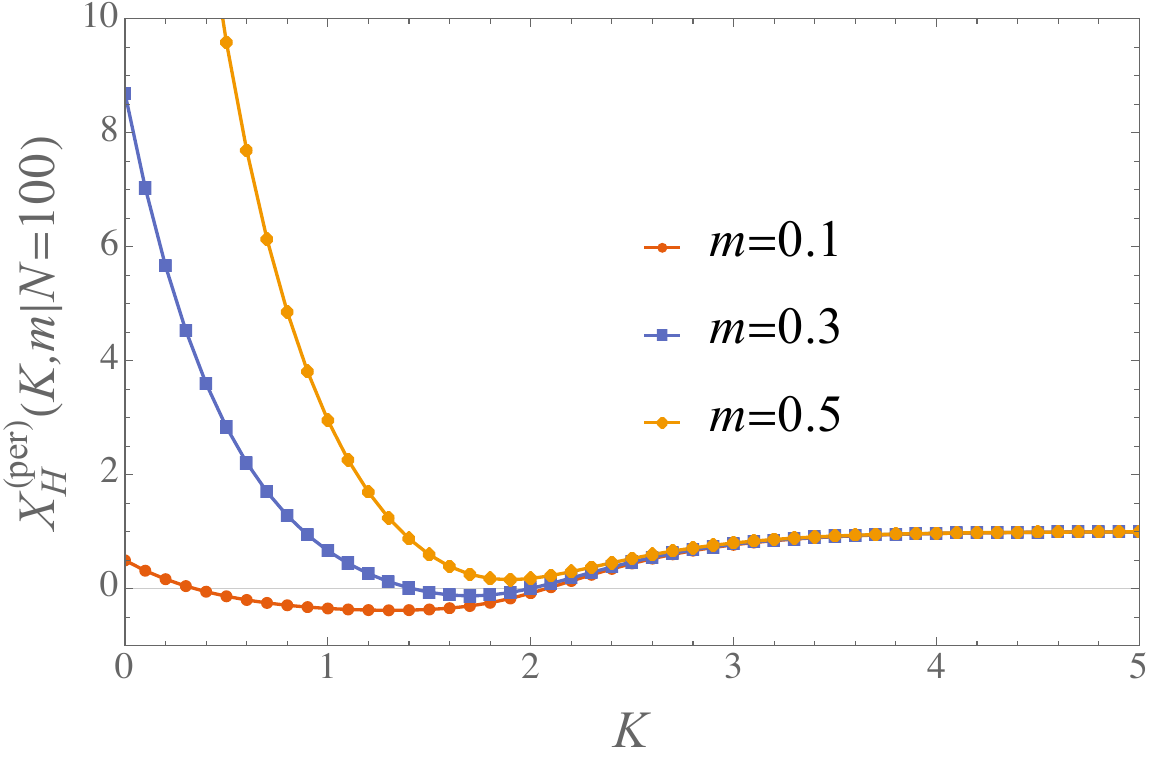}
	\caption{The behavior of the  function $X_{\rm H}^{(\rm per)}(K,m|N)$ (see \eq{eq:figeq3}) with $N=100$ and for $m=0.1, 0.3$, and $m=0.5$. We observe that the function is \textit{positive}  for large values of $K$ and  \textit{negative} for relatively small values of $K$ provided $m$ is also relatively small. For large $m$ the force is always repulsive, irrespective on the value of $K$. The same is also true for very small values of $K$, independent on the values of $m$.  The logarithmic behavior of the free energy of the finite Ising chain with periodic boundary conditions noted in item 1 of the comments above lead to the limit   $X_{\rm H}^{(\rm per)}(K\to\infty,m|N)=1$.  }
	\label{fig:Helmholtz}
\end{figure}

\begin{figure}[htbp]
	\includegraphics[width=\columnwidth]{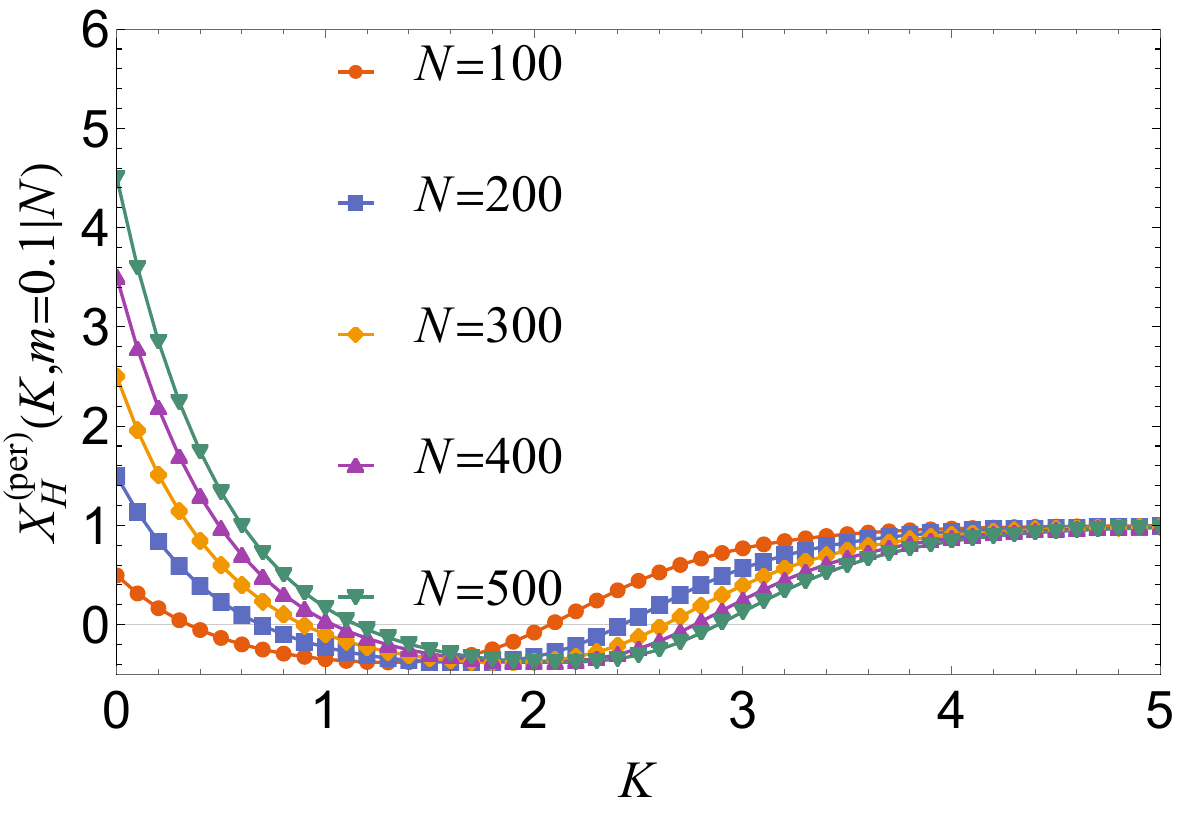}
	\caption{The behavior of the function $X_{\rm H}^{(\rm per)}(K,m|N)$ (see \eq{eq:figeq3}) with $N=100,200, 300, 400$ and $N=500$. We observe that the function is \textit{positive}  for large and  for small enough values of $K$, while being \textit{negative} for relatively moderate  values of $K$, \textit{irrespective} of the value of $N$. The larger $N$, the stronger the repulsion  is for a small enough $K$; the force in the latter regime is strongly repulsive, irrespective on the value of $N$.   }
	\label{fig:Helmholtz2}
\end{figure}

\begin{figure}[htbp]
	\includegraphics[width=\columnwidth]{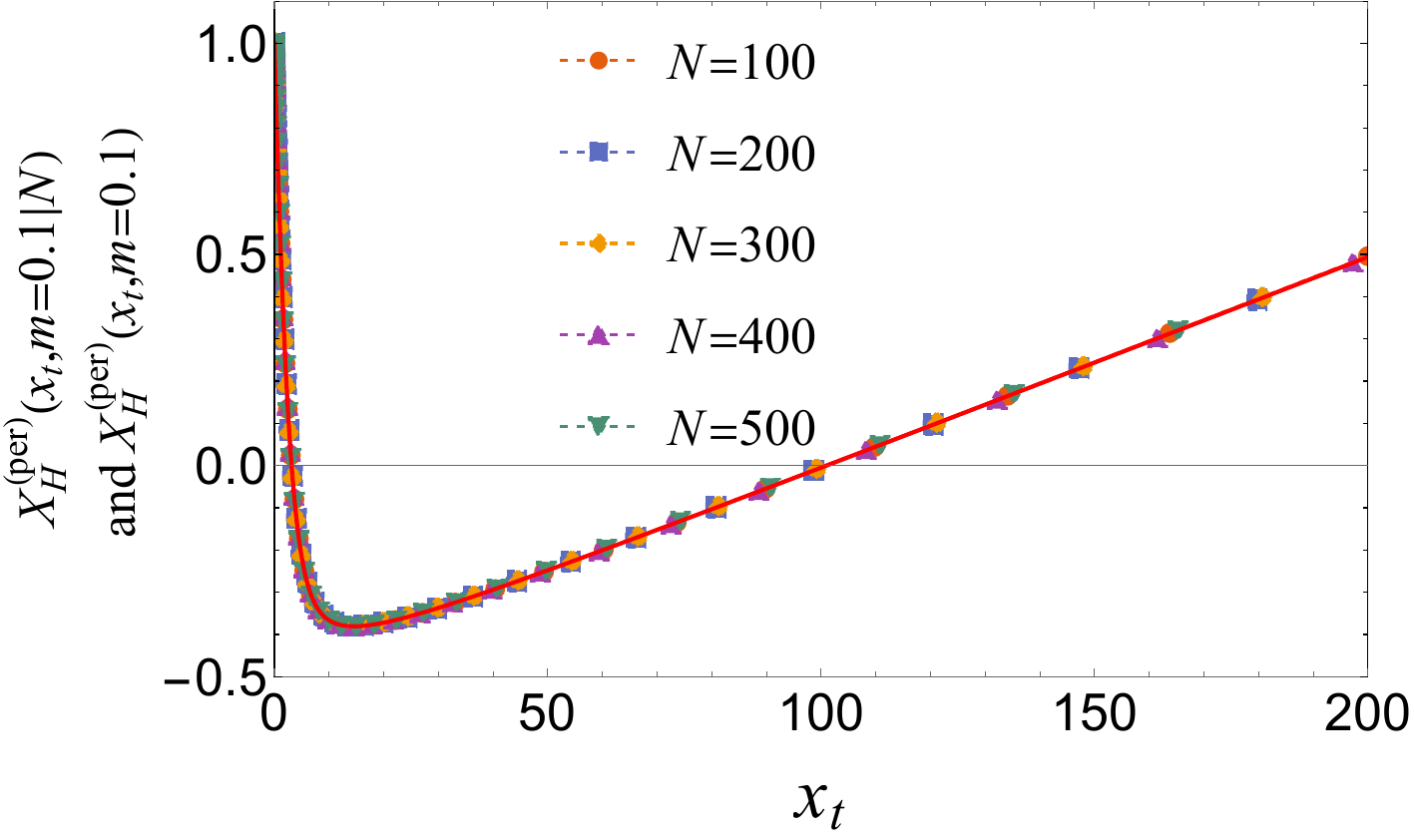}
	\caption{The behavior of the scaling function $X_{\rm H}^{(\rm per)}(x_t,m)$ for $m=0.1$. The inspection of the results obtained numerically from  \eq{eq:statistical-sum} with $N=100,200, 300, 400$ and $N=500$, and that one from \eq{eq:scalingform} demonstrate perfect scaling and agreement between each other. We observe that the function is \textit{positive}  for large values of $x_t$, \textit{negative} for relatively moderate values of $x_t$, and again strongly repulsive for small values of $x_t$. }
	\label{fig:Helmholtz3}
\end{figure}

Note that $m$ can also be seen as a sort of generalized ``charge,'' or symmetry value, which is conserved both inside and outside the system. Given the free energy derivable from the partition function, one is in a position to determine the fluctuation-induced Helmholtz force on a finite Ising chain in contact with a ``bulk,'' chain of infinite extent. The results of such a calculation are shown in Figs. \ref{fig:Helmholtz}---\ref{fig:Helmholtz3}. Along the lines of \eq{HelmDef} and \eq{excess_free_energy_definition_M},  in terms of notations used specifically for the Ising chain, the force is minus the derivative with respect to $N$ of the combined Helmholtz free energy
\begin{equation}
\mathcal{F} = -\ln\left( \mathcal{Z}^{({\rm per})}(N,K,M)\right) + (\mathcal{N}-N)F_H(K,m). \label{eq:figeq1}
\end{equation}
Here $F_H$ is the Helmholtz free energy density of a ``bulk'' neighboring Ising chain. The term proportional to $\mathcal{N}$ can be ignored as a background contribution to the overall free energy. The quantities  $M$, $m$ and $K$ are kept constant in the process of differentiation, after which $M$ is set equal to $mN$. This yields the fluctuation induced Helmholtz force $F_H^{({\rm per})}(K,m,N)$.
Multiplying the result for $F_H^{({\rm per})}(K,m,N)$ by $N$ provides the function $X_H^{({\rm per})}(K,m|N)$
\begin{equation}
X_H^{({\rm per})}(K,m|N)=N	F_H^{({\rm per})}(K,m,N) \label{eq:figeq3}.
\end{equation} 
Its behavior is shown in Figs. \ref{fig:Helmholtz} and \ref{fig:Helmholtz2}. Fig \ref{fig:Helmholtz} shows its behavior as a function of $K$ for $N=100$, and $m=0.1, 0.3$ and $m=0.5$, while Fig. \ref{fig:Helmholtz2} shows it for $m=0.1$ and $N=100, 200, 300, 400$, and $N=500$. Focusing on the scaling regime ($K$ and $N$ both large compared to 1) we end up with the $N$-independent scaling function $X^{({\rm per})}(x_t,m)$. 
Figure \ref{fig:Helmholtz3} shows the behavior of this quantity as a function of $x_t$ for $m=0.1$.

The plots in Fig. \ref{fig:Helmholtz} show that the fluctuation induced force studied has a behavior that is similar to one appearing in some versions of the Big Bang theory---strong repulsion at high temperatures, transitioning to moderate attraction for intermediate values of the temperature, and then back to repulsion, albeit much weaker than during the initial period of highest temperature \cite{inflationary}.  

Our text demonstrates that one can define \textit{ensemble-dependent} fluctuation-induced forces and study their behaviors.  To the best of our knowledge, this issue has not been explored up to now. It is worth noting that, as in the studied case of Helmholz's forces, their behaviors can be quite different from the well-known one of the Casimir forces. 

\section*{Acknowledgements}
DD gratefully acknowledges the discussions and exchange of information with Prof. Siegfried Dietrich and Dr. Markus Gross on some aspects of the current work. We are also grateful to the authors of References \cite{Henkel} and \cite{Racz} for calling our attention to their work. DD acknowledges the financial support by Grant No
BG05M2OP001-1.002-0011-C02 financed by OP SESG 2014-2020 and by the EU
through the ESIFs.


\begin{appendix}

	\section{Antiperiodic and Dirichlet boundary conditions}

Here we present some details on the derivation of the partition function of the Ising chain with fixed magnetization $M$ in the case of antiperiodic and Dirichlet boundary conditions. The final results for these boundary conditions are reported in the main text. 

\subsection{The case of antiperiodic boundary conditions}

The only difference between antiperiodic and periodic boundary conditions is the existence of a``wrong'' bond. This introduces the factor $e^{-2K}$, unless the bond is at a domain boundary, in which case we attach the factor 1. The first factor of $N$ in Eq. (\ref{eq:stat-sum-expansion}) in the article becomes $e^{-2K}(N-8)+8e^{2K}$, where 8 is the number of domain walls. The end-result of this calculation is Eq. (\ref{eq:cf33}) in the article.

\subsection{The case of Dirichlet boundary conditions} 

\label{DD-bc}

In order to derive Eq.(\ref{eq:DDbc-main-text-appendix}) in the article, we have to consider two cases. In the first case there is an odd number of domain walls, and in the second case there is an even number of domain walls. Let's start with the first case.

\subsubsection{Odd number of domain  walls}

First, we'll look at the simplest instance: a one-dimensional Ising model with Dirichlet boundary conditions and a single domain wall. This domain wall separates two domains: a spin-up domain with $N_+$ spins and a spin-down domain with $N_-$ spins. Given Eqs. (2.2), (2.3) in the article, we can express those two numbers in terms of $N$ and $M$. For the time being we will write expressions in terms of the number of up and down spins in the system. There are precisely two ways to divide the system into two domains: with the up-spin domain to the left or the down-spin to the left. This means that the combinatorial factor here is 2. The contribution to the Boltzmann factor associated with this arrangement is $e^{-2K}$, the cost of a single domain wall. This multiplies the Boltzmann factor $e^{(N-1)K}$ - the Boltzmann factor of a fully aligned system in which the leftmost spin does not interact with the rightmost spin, as would be the case if the system were periodically continued. 

\begin{figure}[htbp]
	\begin{center}
		\includegraphics[width=3in]{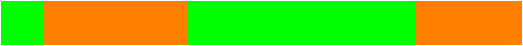}
		\caption{Three domain walls and four domains of alternatively up and down spins}
		\label{fig:fourdomains}
	\end{center}
\end{figure}

Next, what if there are three domain walls? In this case, the system splits into four domains, which alternate, as shown in Fig. \ref{fig:fourdomains}. 

We can calculate the number of ways in which we can split the up and down spins into two domains. Following the argument in Section \textbf{II} of the main text, we find the factor
\begin{equation}
	2 \times (N_+-1)(N_--1).   \label{eq:cf11}
\end{equation}
Expressing $N_+$ and $N_-$ in terms of $N$ and $M$, we end up with
\begin{equation}
	2 \times \frac{(N-2)^2-M^2}{4}. \label{eq:cf12}
\end{equation}
This expression holds only if the terms in the denominator are positive. The factor of 2 in front expresses the fact that the leftmost domain can consist of up spins or of down spins. The contribution to the Boltzmann factor is $e^{-6K}$, the cost of three domain walls. Similarly, if there are five domain walls, then there are six domains, and the number of ways of dividing the up spins into three domains and the down spins into three domains is 
\begin{eqnarray}
	\lefteqn{2 \times \frac{ (N_+-1)(N_+-2)}{2!} \times \frac{(N_--1)(N_--2)}{2!}} \nonumber \\
	& = & 2 \times \frac{((N-2)^2-M^2)((N-4)^2-M^2)}{(2!)^2 4^2}. \label{eq:cf13}
\end{eqnarray}
Once again, the overall factor of 2 reflects the fact that the leftmost domain can be either of up spins or down spins. This is all with the proviso that all factors in the numerator are positive. The contribution to the Boltzmann factor is $e^{-10K}$. 

This can clearly be systemized.

\subsubsection{Even number of domain walls.}

\begin{figure}[htbp]
	\begin{center}
		\includegraphics[width=3in]{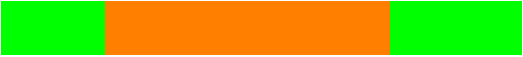}
		\caption{Two domain walls and three domains of alternatively up and down spins.}
		\label{fig:threedomains}
	\end{center}
\end{figure}

An even number of domain walls corresponds to an odd number of domains. In the simplest non-trivial case there are two walls and three domains, as shown in Fig. \ref{fig:threedomains}.
Suppose that the domains are up-spins, down-spins, then up-spins. We have to figure out how many ways there are of arranging this. The combinatorial factor is 
\begin{equation}
	N_+-1 = \frac{N-2+M}{2}, \label{eq:cf14}
\end{equation}
where the expression only applies if it is greater than zero. The contribution to the Boltzmann factor is $e^{-4K}$. The other possibility is that the domains are down-spins, up-spins then down-spins again. In this case the combinatorial factor is 
\begin{equation}
	N_--1 = \frac{N-2-M}{2} \label{eq:cf15}
\end{equation} 
with the same Boltzmann factor, $e^{-4K}$ and the same proviso that the expression must be positive. 

The next possibility is four domain walls, resulting in five alternating domains. If there are three up-spin domains and two down-spin domains then the combinatorial factor is 
\begin{eqnarray}
	\lefteqn{\frac{(N_+-1)(N_+-2)}{2!} \times (N_--1)} \nonumber \\ & =& \frac{(N-2+M)(N-4+M)}{4 \times 2!} \times \frac{N-2-M}{2}. \label{eq:cf16}
\end{eqnarray} 
The Boltzmann factor here is $e^{-8K}$. If there are three down-spin domains and two up-spin domains, the combinatorial factor is
\begin{eqnarray}
	\lefteqn{\frac{(N_--1)(N_--2)}{2!} \times (N_+-1) } \nonumber \\ & = & \frac{(N-2-M)(N-4-M)}{4 \times 2!} \times \frac{N-2+M}{2} \label{eq:cf17}
\end{eqnarray}
with the same Boltzmann factor. In both cases above, as in all cases, each factor in the numerator must be greater than zero. 

Again, systemization is straightforward.

\subsubsection{The maximum number of domain walls for fixed values of $N$ and $M$: odd number of domain walls}
Here, the number of domain walls is $2k+1$. We will proceed under the assumption that $M$ is positive. In that case, we look for the largest value of the even number $2k$ such that $N-2k-M>0$. Solving this inequality for $k$, we find 
\begin{equation}
	k<\frac{N-M}{2}. \label{eq:cf18}
\end{equation}
The thing to remember is that $k=0$ is an acceptable possibility. This corresponds to the case of a single domain wall. 

\subsubsection{The maximum number of domain walls for fixed values of $N$ and $M$: even number of domain walls}

In this case there are $2k$ domain walls. Here the situation is a bit more complicated, as we have to deal with two possibilities. The first is that the number of up-spin domains is one larger than the number of down-spin domains, and the second is that the number of down-spin domains is one greater than the number of up-spin domains. We will address each case separately

\vspace{0.1in}

\noindent { \textit{ * One more up-spin domain}}

\vspace{0.1in}

In this case, we seek the largest value of the integer $k$ such that the following two inequalities hold:
\begin{eqnarray}
	N+M-2k>0 \label{eq:cf19}  \\
	N-M-2(k-1) >0. \label{eq:cf20}
\end{eqnarray}
These two inequalities lead to the two equivalent inequalities
\begin{eqnarray}
	k<\frac{N+M}{2} \label{eq:cf21}  \\
	k<\frac{N-M}{2} +1 \label{eq:cf22}.
\end{eqnarray}
Comparing the right hand sides of (\ref{eq:cf21}) and (\ref{eq:cf22}), we find that the operant inequalities are
\begin{equation}
	k< \left\{ \begin{array}{ll} \frac{N+M}{2} & M=0 \\ \frac{N-M}{2} +1 & M >0 \end{array}  \right. \label{eq:cf23}
\end{equation}

\vspace{0.1in}

{ \textit{* One more up-spin domain}}

\vspace{0.1in}

In this case, we find that the requirement on $k$ is always
\begin{equation}
	K< \frac{N-M}{2}.  \label{eq:cf24}
\end{equation}

\subsubsection{Explicit representation of the partition function for the one-dimensional Ising model with Dirichlet boundary conditions}

Given the following definitions
\begin{eqnarray}
	\lefteqn{P_{e+}(N,M,j)} \nonumber \\ & = & \binom{(N-M)/2-1}{j-1} \binom{(M+N)/2-1}{j}, \label{eq:cf28} \\
	\lefteqn{P_{e-}(N,M,j)} \nonumber \\ & = & \binom{(N-M)/2-1}{j} \binom{(M+N)/2-1}{j-1}, \label{eq:cf29} \\
	\lefteqn{P_o(N,M,j) } \nonumber \\ & = & 2 \binom{(N-M)/2-1}{j} \binom{(M+N)/2-1}{j}, \label{eq:cf30}
\end{eqnarray}
the partition function for the one-dimensional Ising model with $N$ spins and a total magnetization $M$ with coupling constant $K$ subject to Dirichlet boundary conditions can be written in the following form
\begin{eqnarray}
	\lefteqn{Z^{(D)}(N,M,K)} \nonumber \\ & =& e^{(N-1) K} \nonumber \\ &&\Bigg\{ \sum_{j=1}^{\infty} e^{-4 j K}\Big[ P_{e+}(N,M,j)  +P_{e-}(N,M,j)\Big] \nonumber \\ && +\sum_{j=0}^{\infty}  e^{-2 (2 j+1) K}  P_o(N,M,j)\Bigg\} \nonumber \\ \label{eq:cf31}
\end{eqnarray}
The upper limit on the sums is automatically satisfied for integer values of $N$ and $M$ as long as both are even or odd, which correspond to the allowed values of those two quantities. Otherwise, the expression on the right hand side of (\ref{eq:cf31}) can be taken to be an interpolation formula. The sums can be evaluated in terms of hypergeometric functions. The result of the evaluation  for $|M|<N$ is Eq. (\ref{eq:DDbc-main-text-appendix}) in the main text. 

\end{appendix}

\end{document}